\documentclass[aps,prl,twocolumn,letter]{revtex4-1}

\usepackage{graphicx}  
\usepackage{bm}        
\usepackage{amssymb}   
\usepackage{color}
\usepackage{ulem}
\usepackage{bbding}
\usepackage{ifsym}
\usepackage{MnSymbol}
\usepackage{pifont}
\usepackage{latexsym}
\usepackage{wasysym}
\usepackage{hyperref}
\hypersetup{
    colorlinks=true,
    linkcolor=blue,
    anchorcolor=blue,
    citecolor=blue,
}

\begin{document}

\title{Universality in quasi-2D granular shock fronts above an intruder}
\date{\today}
\author{M. Yasinul Karim, Eric I. Corwin}
\affiliation{Materials Science Institute and Department of Physics, University of Oregon, Eugene, Oregon 97403}

\begin{abstract}

We experimentally study quasi-2d dilute granular flow around intruders whose shape, size and relative impact speed are systematically varied. Direct measurement of the flow field reveals that three in-principle independent measurements of the non-uniformity of the flow field are in fact all linearly related: 1) granular temperature, 2) flow field divergence and 3) shear-strain rate.  The shock front is defined as the local maxima in each of these measurements. The shape of the shock front is well described by an inverted catenary and is driven by the formation of a dynamic arch during steady flow. We find universality in the functional form of the shock front within the range of experimental values probed. Changing the intruder size, concavity and impact speed only results in a scaling and shifting of the shock front. We independently measure the horizontal lift force on the intruder and find that it can be understood as a result of the interplay between the shock profile and the intruder shape.

\end{abstract}

\maketitle

Lift forces \cite{soller_drag_2006, ding_drag_2011,  potiguar_lift_2011, potiguar_lift_2013} on and shock formation \cite{haff_grain_1983, rericha_shocks_2001, amarouchene_speed_2006} around intruders in granular flows are well known phenomena. Rericha \textit{et al.} \cite{rericha_shocks_2001} and Boudet \textit{et al.}  \cite{boudet_shock_2008,  boudet_drag_2010} have demonstrated that shocks analogous to those in fluid flows are formed around symmetrical intruders in dilute granular flows. Numerical studies by Potiguar \cite{potiguar_lift_2011} have shown that in the case of asymmetrical intruders the drag-induced lift forces are lower than expected due to the formation of shock waves that act as a shield. However, the relationship between the granular shock front and lift (forces perpendicular to the direction of flow) on an intruder has not been studied in detail. 

\begin{figure}
\includegraphics[width=1\linewidth]{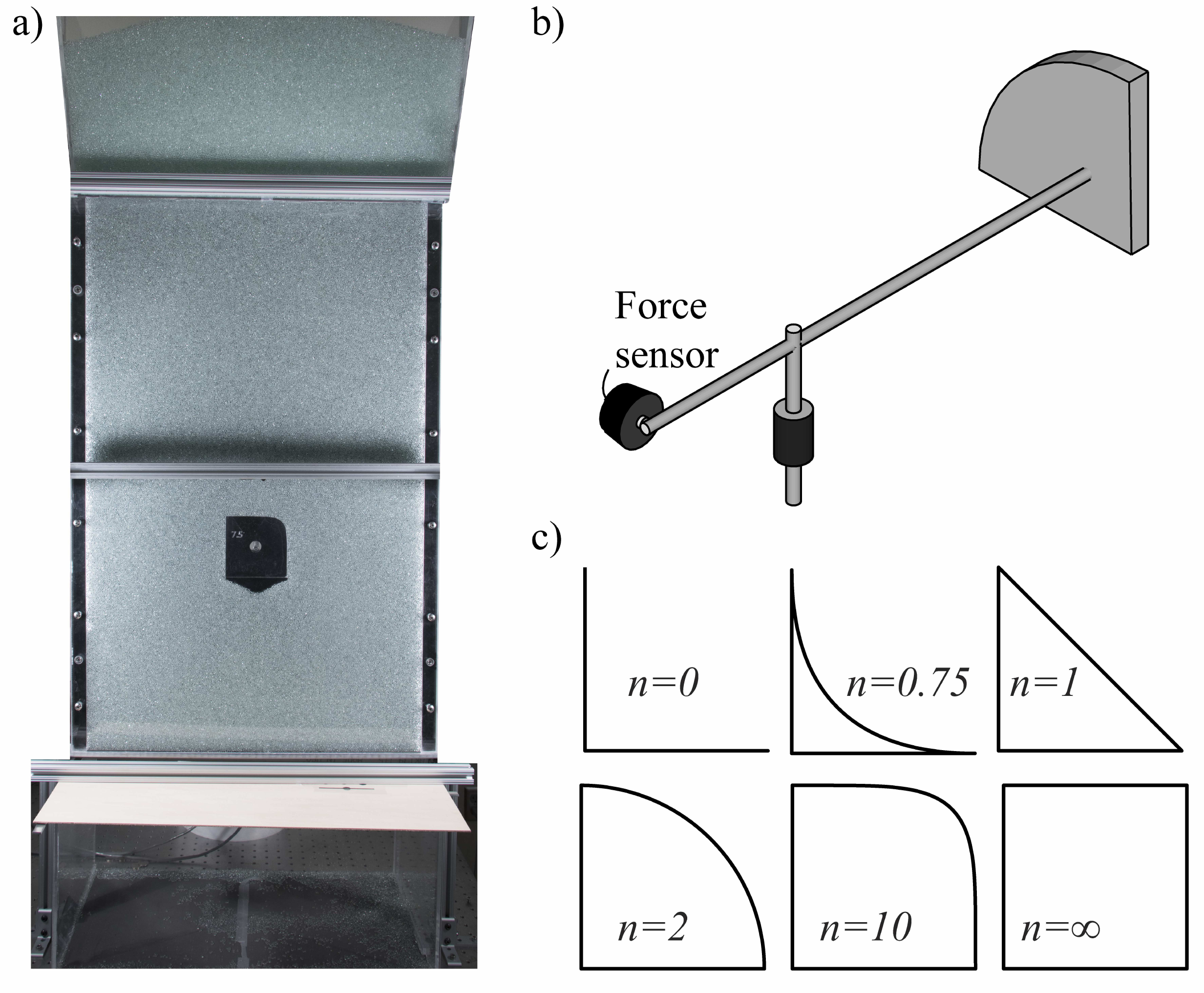}
\caption{a) Front view of experimental setup.  The intruder in the picture corresponds to a super-disk exponent $n=7.5$.  b) Intruder-sensor system schematic. c) Examples of intruders with different super-disk exponents. As $n$ increases the shapes become more convex with $n=\infty$ being a square.}
\label{schematic}
\end{figure}

\begin{figure*}
\includegraphics[width=1\linewidth]{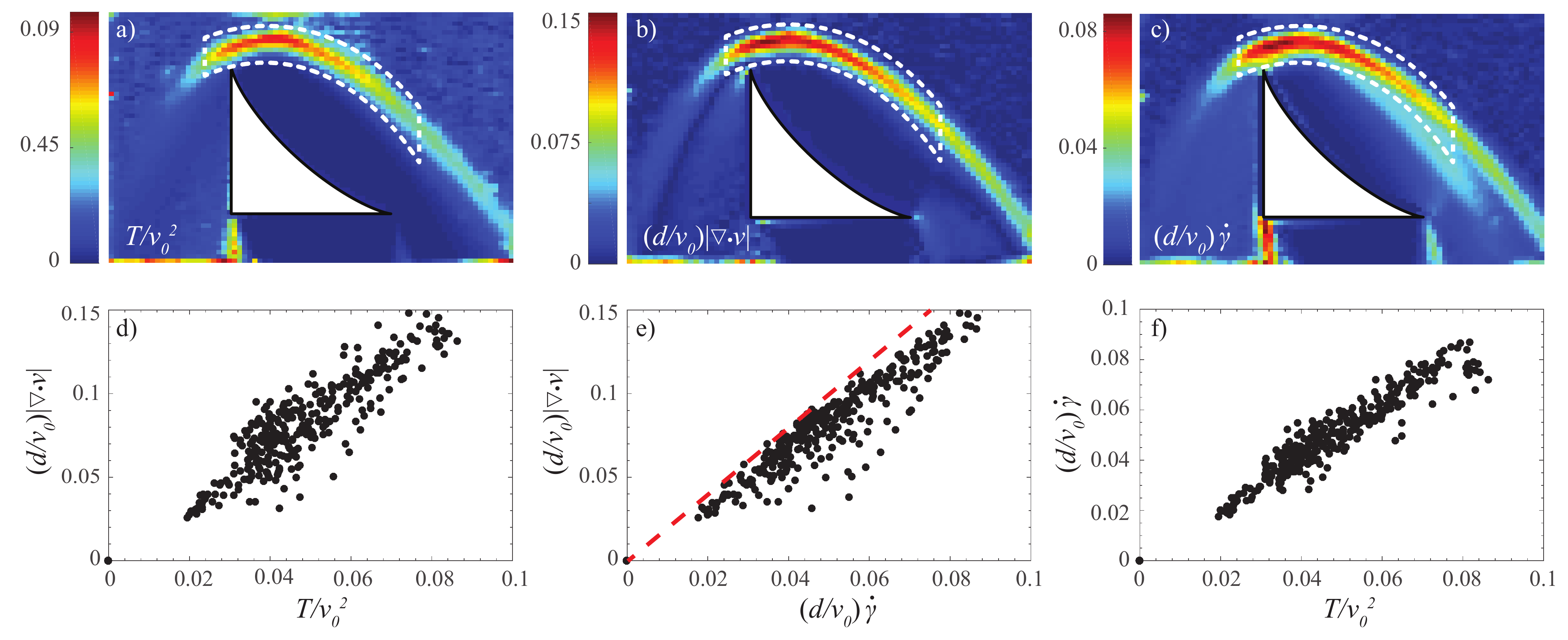}
\caption{a-c) Intensity plot of scaled granular temperature, absolute divergence of the flow field, and local shear-strain rate for $n=0.75$. d - f) Scatter plots of scaled flow-field divergence versus temperature, flow field divergence versus local shear rate, and shear rate versus temperature respectively, measured from all points enclosed by the white dotted regions.}
\label{mapPanel}
\end{figure*}

In this letter we systematically vary the shape of an intruder in a dilute granular flow and measure the resulting grain flow fields and lift forces.  We find that the position of the shock front is well-described by each of the coordinates of the local maxima of granular temperature, divergence, and local shear-strain rate, which are all found to be linearly related.  The shock front maintains a symmetric shape and its center point closely tracks the center of mass of the intruder.  We find that the asymmetry of the intruder alters the shock boundary and this in turn alters the lift force on the intruder. We propose a model for calculating lift forces which agrees with experimental data. We demonstrate that the force in the direction perpendicular to flow is the result of three mechanisms: 1) static loading from the dead zone above the intruder, 2) impact forces from the freely falling grains hitting the shock front, and 3) net momentum transfer due to mass ejection from the dead zone.

Our experimental setup consists of a rectangular quasi-2D granular hopper that is 1.25 cm deep, 55.8 cm wide, and 90 cm in height.  The hopper is constructed from 1/4 inch acrylic as shown in Figure \ref{schematic}a. The front of the hopper is clear and the back is opaque to allow for quantitative imaging of the flow. The hopper is fed by a reservoir of 3 mm glass beads (Mo-Sci Corp) to provide a steady flow rate of approximately 400 particles cm$^{-2}$s$^{-1}$. A laser-cut acrylic intruder is placed inside the hopper and is constructed to be 1.1 cm thick, slightly thinner than the hopper cavity to prevent frictional contact with the front and back walls.  The intruder is typically mounted at $h$=57 cm below the reservoir, where the bead velocity in the absence of the intruder would be $v_{0}$=3.3 m/s, comparable to the flow speeds in \cite{rericha_shocks_2001, amarouchene_speed_2006, clark_steady_2016}. Changing $h$ corresponds to changing the impact speed $v_{0}$ of the grains; $v_{0}\propto \sqrt{h}$.

Figure \ref{schematic}b shows the intruder-sensor system. The intruder is connected to one end of a 30 cm aluminum rod. The other end of the rod is held in place by a force sensor to measure lateral forces (Measurement Specialties FC22) and a pre-load which is subtracted from the measurements. This is fixed to a second rod to form a T-shaped structure which is held by low friction bearings.

The intruder shape is defined as the first quadrant of the superdisk equation \cite{jiao_optimal_2009} 
\begin{equation} 
y = (1 - x^n)^{\frac{1}{n}},
\label{intruderEq}
\end{equation}
where $x$ and $y$ are the horizontal and vertical coordinates. The exponent $n$ controls the concavity/symmetry of the shape as illustrated by the six shapes in Figure \ref{schematic}c.  The superdisk equation provides a family of shapes interpolating from the asymmetric ``L" at $n=0$ to a triangle at $n=1$, a quarter circle at $n=2$ and a square at $n=\infty$.  Unless otherwise specified the length $R$ of the straight edge is held constant at 4 inches.   

At the beginning of each measurement the bottom of the hopper is blocked off and it is filled with beads. When the beads are released the force sensor and a high-speed camera (Phantom M310) start recording. We ignore the initial transient (approximately 0-4 seconds) when beads begin to flow and keep only the measurements taken in steady state - when the flow rate is roughly constant.  After about 10 seconds the reservoir is depleted.

For each value $n$ we measure lateral forces on, and granular flow field around the intruder. Bead velocities are calculated using particle image velocimetry \cite{thielicke_pivlab_2014}. In agreement with previous studies \cite{amarouchene_dynamic_2001, wassgren_dilute_2003, knoll_numerical_2007, boudet_shock_2008, meyer_transient_2013, vilquin_structure_2016} in steady state flow we observe a nearly stationary, densely packed pile of beads with a shock front above the leading edge of the intruder. This shock front can be identified by three different measurements - the local maxima of granular temperature $T$, flow divergence $f_d$, and shear strain rate $\dot\gamma$: 
\begin{align} 
T &= \langle \vec{v}\cdot \vec{v} \rangle - \langle\vec{v} \rangle \cdot \langle\vec{v} \rangle, \label{gTemp}\\ 
f_d &= | \nabla\cdot\vec{v}|, \textrm{and} \label{flowDiv}\\ 
\dot{\gamma} &= (d_{1}-d_{2})/2 \label{shearRate}, 
\end{align}
where $\vec{v}$ is the velocity at any given point. Angled brackets denote time averages and $d_{1}$ and $d_{2}$ are eigenvalues of the strain rate tensor $\textbf{D}=\nabla \vec{v}+(\nabla \vec{v})^{\textbf{T}}$, following the method described by Clark \textit{et al.} \cite{clark_steady_2016}. 

For all values of $n$, these three tests pick out the same feature above the intruder. They are shown for a representative intruder with $n=0.75$ in Figure \ref{mapPanel}a-c.  These quantities are all close to zero everywhere else in the flow field. By picking out only the points inside the same dotted regions in Figures \ref{mapPanel}a-c, we plot divergence versus temperature, divergence versus shear rate and shear rate versus temperature in Figures \ref{mapPanel}d-f respectively. We non-dimensionalize these quantities by scaling them as $T/v_{0}^2$, $(d/v_{0}) | \nabla \cdot \vec{v}|$, and $(d/v_{0})\dot \gamma$, where $v_{0}$ is the theoretical free-fall speed of beads at the intruder's center of mass and $d$ is the bead diameter.  In Figure \ref{mapPanel}e the red dotted line corresponds to a slope of 2. Since $|f_{d}|=|d_{1}+d_{2}|$ and $\dot{\gamma}=d_{1}-d_{2}$ where $d_{1}$ and $d_{2}$ are eigenvalues of the shear strain tensor, \ref{mapPanel}e shows that $|d_{1}+d_{2}|>\frac{1}{2}(d_{1}-d_{2})$. The points in the scatter plot fall below the slope $=2$ line (red dotted line in Figure \ref{mapPanel}e) showing that $d_{1} >> d_{2}$. 

The non-zero divergence band is a region of high compressibility - grains suddenly decelerate upon impact and some get absorbed into the quasi-static pile on the intruder. The same band also displays high variance in the velocity field due to sudden grain-grain collisions and consequent velocity changes. The linear relationship between the three measured quantities are unexpected and remain to be studied further in a future study.    

\begin{figure}
\includegraphics[width=1\linewidth]{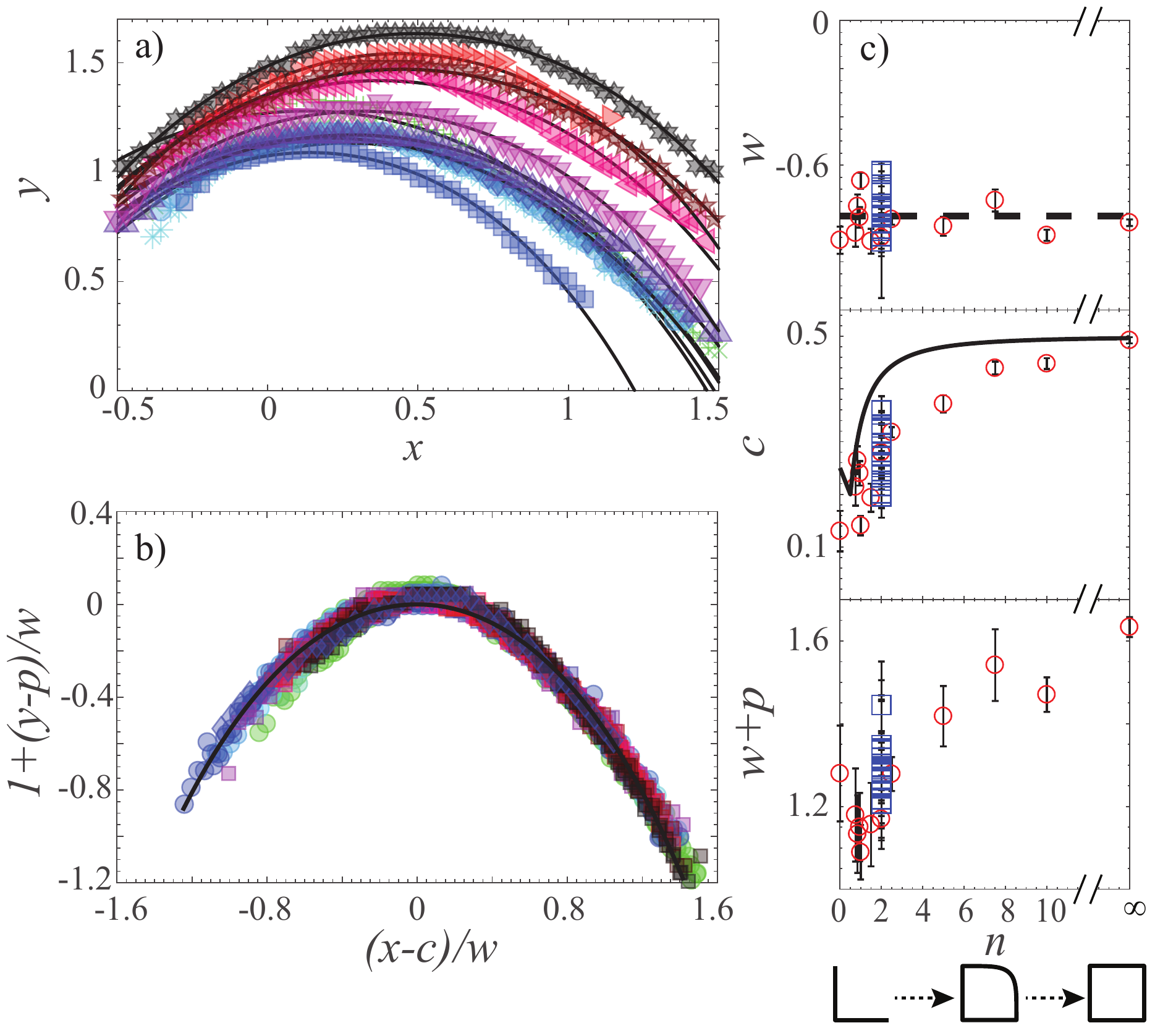}
\caption{a) Plots of shock fronts. Scaled $x$ and $y$ coordinates are in units of $R$, the vertical length of the intruder. The data is shown for $n=0$ ($\plus$), $0.75$ ($\times$), $0.85$ ($\ast$), $0.95$ ($\circ$), $1$ ($\Box$), $1.5$ ($\Diamond$), $2$ ($\triangle$), $2.5$ ($\triangledown$), $5$ ($\triangleleft$), $7.5$ ($\triangleright$), $10$ ($\largestar$), $\infty$ ($\davidsstar$). Solid black lines are fits to Equation \ref{modBrac}. b) Plot of scaled shock boundaries for all probed values of $n$ with constant $R=4$ inches shown in circles. Squares represent all tested intruder locations $h\in\{ 23, 34, 46, 57, 69, 84 \}$cm and intruder radii, $R\in\{2,3,4,7\}$inches for constant $n=$2. Diamond symbols represent data from an asymmetric intruder with random features on it (shown in Figure 2 of the Supplement). The black curve is the inverted catenary given by Equation \ref{modBrac}. c) Top to bottom - plots of the fit parameters $w$, $c$ and $w+p$ as functions of $n$ respectively. Red circles represent data from varying $n$ with constant $R$ and blue squares represent data from fixed $n=$2 but different combinations of $R$ and impact speed $v_{0}$. The dashed line shows the mean width $|\bar{w}| \approx$0.8. In the $c$ versus $n$ plot the dark solid curve shows center of mass of the intruders shift with varying $n$. }
\label{curveFitPanel}
\end{figure}

Figure \ref{curveFitPanel}a shows the extracted shock front for intruders of varying $n$. The plots show that the shock front maintains the same functional form as the intruder geometry is varied, with the curves shifting right and increasing in peak height with increasing $n$. For any given obstacle shape we find that the volume of trapped beads remains constant. Volume fluctuations are low so we assume a system with negligible dilatancy and bounded by a shock front that prefers to maintain a shape dictated by force constraints.   

\begin{figure}
\includegraphics[width=1\linewidth]{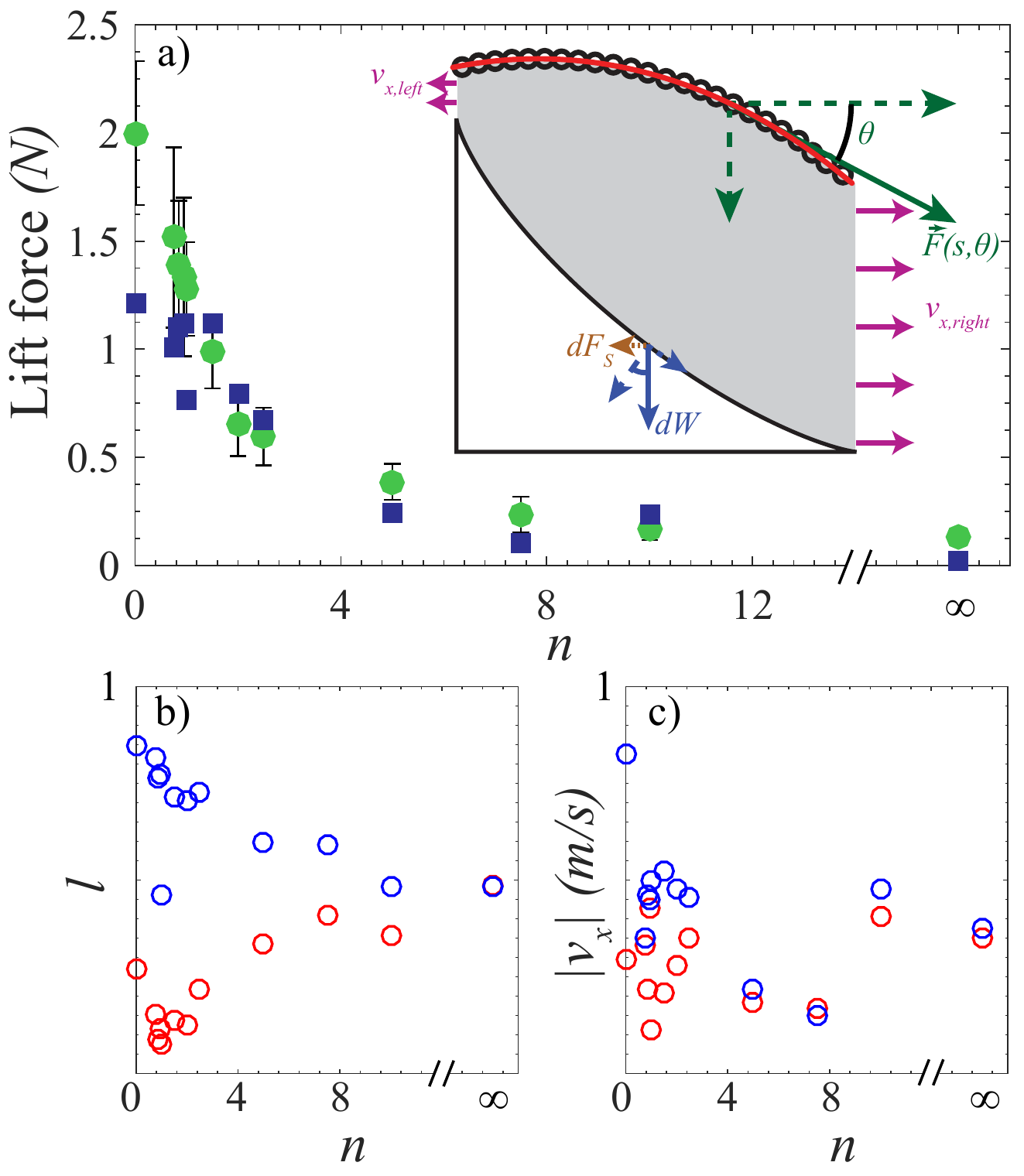}
\caption{a) Plot of horizontal lift forces from force sensor (green circle) and calculated lift forces (dark blue squares) as a function of $n$. Lift is maximum for $n\rightarrow 0$ and decreases to $0$ as $n\rightarrow\infty$. Inset is an illustration showing the intruder, outlined in black, trapped grains, in grey and shock profile in red. The solid green arrow represents the tangential net force $\vec{F}(s,\theta)$ on a small segment of the shock front. Dotted green arrows are components of this vector. Purple arrows labeled $v_{x,0/1}$ represent horizontal ejected grain velocities from the trapped region. b) Plot of the left (red) and right (blue) exit heights $l$ as a function of $n$. c) Plot of absolute horizontal left (red) and right (blue) exit velocities $|v_{x}|$ versus $n$.}
\label{forcePlot}
\end{figure}

During steady-state flow the shock front must remain stable. Any vertical or horizontal net force will result in a time variant shock front. To enforce stability of the arch-like shock front given the constraints of gravity the net force must be tangential to the surface of the shock as illustrated by the green arrow in Figure \ref{forcePlot}a (inset). This net force ensures that particles not absorbed by the quasi-static pile are cleared away from it such that the shock front remains stable. The ejection of particles from the left and right, illustrated by purple arrows in Figure \ref{forcePlot}a (inset), also ensure a stable shock position during steady flow.

Force balance also implies that the $i$th particle in the quasi-static layer is held in place at an angle of repose $\theta_{i}$ such that $K_{i}\mu = \tan{\theta_{i}}$, where $\mu$ is coefficient of friction and $K_{i}$ is a friction mobilization factor at the $i$th location. Thus $\tan{\theta_{i}} = \frac{\Delta y_{i}}{\Delta x_{i}}$. 

We define $s_{i}$ as the arc length spanning the horiontal distance $l$, the particle diameter. For a convex shock profile the ratio $s_{i}/l$ is an approximate measure of $K_{i}$. A higher ratio implies more friction mobilized to constrain the particle. When the tangent to $s_{i}$ is horizontal, $\tan\theta_{i} = 0$ and $s_{i}/l=1$. To enforce this condition we assume that to first order $K_{i} \propto \left(s_{i}/l - 1 \right)$. In the discrete limit, for some constant of proportionality $C$,
\begin{equation}
\left( \frac{s_{i}}{l} - 1 \right)C\mu = \tan{\theta_{i}}=\frac{\Delta y_{i}}{\Delta x_{i}}.
\label{discrete} 
\end{equation}

In the continuum limit, taking the $x$-derivative on both sides of Equation \ref{discrete}:
\begin{equation}
(C \mu/l)\frac{ds}{dx}=\frac{1}{w}\frac{ds}{dx} = \frac{d^{2}y}{dx^2}.
\label{continuum}
\end{equation}
Equation \ref{continuum} has the solution  
\begin{equation} 
y=w\cosh \left( \frac{x-c}{w} \right) +p.
\label{modBrac}
\end{equation}
Equation \ref{modBrac} is an inverted catenary where $|w|$ is the characteristic width of the hyperbolic cosine curve, $c$ sets the center of the curve, and $p$ is the vertical offset such that $y(x=c)=w+p$. Surprisingly this description of the shock front has the same solution as the shape of an arch in a jammed pipe \cite{mounfield_theoretical_1996}.  

The measured shock fronts are well fit by Equation \ref{modBrac} as demonstrated by the dark lines on Figure \ref{curveFitPanel}a. Thus we find that in steady granular flow the shock front (ignoring temporal fluctuations) is a dynamic arch. By scaling out the fit parameters we find excellent collapse onto a single master curve, as demonstrated in figure \ref{curveFitPanel}b. This indicates that the influence of the intruder shape is limited to changing the centerpoint and height of the resulting shock wave. 

To test the effect of intruder size, $R$, for a super-disk exponent $n=2$ we measured several combinations of varying $R$ and height $h$. Figure \ref{curveFitPanel}b includes measurements for all probed values of $n$ and all tested combinations $(R,h)$ for $n=2$. This provides strong evidence for the universality of the shape of the shock front; varying any of the parameters, intruder shape $n$, intruder size $R$ and impact or incident grain velocity $v_{0}$, does not change the shape of the shock front. Finally, we created an intruder with no obvious symmetry (shown in Figure 2 of the Supplement) and found that its shock front, plotted as diamonds in Figure \ref{curveFitPanel}b, also collapses to the same master curve.

We obtain the constants $w$, $c$ and $p$ as fit parameters to Equation \ref{modBrac} and plot them as functions of $n$ in Figure \ref{curveFitPanel}c, where red circles represent data for varying $n$ and blue squares represent data for different combinations $(R,h)$ for $n=2$. The plot of width $w$ versus $n$ demonstrates that the scaled width of the catenary is independent of the shape parameter $n$ and the dotted line denotes the mean value $-0.8$ of the curve width. The plot of center $c$ versus $n$ agrees with the qualitative observation from Figure \ref{curveFitPanel}a - the peaks shift away from the vertical edge as $n$ goes from $0\rightarrow\infty$.  The center naturally asymptotes to $c=0.5$ for $n\rightarrow\infty$ as the shape of the intruder approaches a symmetric square. We find that $c$ roughly tracks the intruder center of mass (black solid curve).  Plotting $y(x=c)$, which is the peak height $w+p$, as a function of $n$ demonstrates that the peaks shift upwards with increasing $n$. Thus, we find that the intruder shape parameter $n$ controls the catenary center and peak.

The shape of the catenary affects how grains are distributed on the intruder and thus, in turn, determines the lift forces $F_{L}$.   We show this by directly measuring $F_L$, the lift force perpendicular to the direction of flow, on the intruder. The measured lift as a function of $n$ is plotted as green circles in Figure \ref{forcePlot}a.  The lift is maximum for $n=0$, and decreases with increasing $n$ and asymptotes to $0$ as $n\rightarrow\infty$, in which case the intruder is perfectly symmetric.  As detailed in the supplement we can model this lift force by considering three contributions: 1) the static load from the quasi-static pile, 2) the momentum flux from freely falling particles hitting the shock front and 3) momentum transfer from particles ejected from either side of the quasi-static pile.  

The first contribution is from the horizontal component $dF_{S}$ of the static load $W$ pushing on the curved upper surface of the intruder. For all intruder shapes there is a trapped pile whose upper boundary is the shock front, which we can approximate as a static load.  Because this load is resting on a curved, asymmetric surface it will impart a non-zero horizontal component $F_{S}$, which is calculated explicitly in the supplement. In the limiting cases $n=0$ or $\infty$ this contribution goes to zero.

The second force contribution is from the momentum flux of freely falling particles impacting the shock front. This force, $F_{C}$ is equal to the incident mass per unit time perpendicular to the shock front multiplied by the velocity change in the horizontal direction. The incident velocities along the shock front are taken from PIV measurements, a representative plot of which is shown for $n=0.75$ in Figure 4 of the Supplement.

The third force contribution $F_{flow}$ is from grains being ejected from the area between the shock front and the intruder profile.  These grains are measured to leave with average speeds $v_{x,left}$ and $v_{x,right}$ illustrated by purple arrows in Figure \ref{forcePlot}a inset. Since we keep the width of the hopper constant the scaled heights $l_{left}$ and $l_{right}$, distances from the intruder to the shock front at $x=0$ (red circles) and $x=1$ (blue circles) respectively, are measures of the exit areas. The distances $l_{left/right}$ are plotted as functions of $n$ in Figure \ref{forcePlot}b and are scaled by the intruder size $R$ as before. We find that the difference in the two exit areas ($|l_{left}-l_{right}|$) decreases with increasing $n$ and becomes equal at $n=\infty$. This implies that for equal exit velocities more grains can escape from the exit area at $x=1$, which in turn means greater momentum flux at this edge.   

The mean exit velocities at $x=0$ (red circles) and $x=1$ (blue circles) obtained from PIV measurements are plotted in Figure \ref{forcePlot}c and we find that the speeds are roughly the same and nearly equal for $n>2$.  Thus, for asymmetrical intruders the disparity in granular exit areas becomes the dominating factor.  At smaller $n$, as the shape becomes more concave and asymmetric, the momentum flux on the right at $x=1$ is greater so $F_{flow}$ is higher. For larger values of $n$ the two areas and speeds are comparable so net momentum flux, $F_{flow}$ approaches zero for $n\rightarrow\infty$. 

Our approach to calculating lift is similar to the granular resistive force theory technique (GRFT) pioneered by Zhang and Goldman \cite{goldman_2014} and further explained by Askari and Kamrin \cite{askari_intrusion_2016}. While GRFT has been effective in predicting and modeling dynamics in friction-dominated granular systems \cite{aguilar_2016} Zhang and Goldman conclude that GRFT may not be applicable when inertial forces dominate \cite{goldman_2014}. Our results demonstrate this is indeed the case in such systems; the inertial forces cannot be ignored when computing net lift for instance. In dilute granular flows further complications arise due to the separation of grains into a solid-like and fluid-like region.  

This experiment reveals that in the region near the granular shock front the mean flow field variance, absolute flow field divergence and shear strain rate are linearly related to each other. The local maxima in each of these measurements provide a robust means of identifying the shock front. The shock front is characterized by a universal functional form that is, unlike the fluid flow analog, invariant with respect to intruder size, shape and impact speeds within probed experimental values. To enforce stability given the constraints of gravity and granular impact forces there must always be a tangential force along the shock front. This leads to an appropriate organization of quasi-static grains around the intruder such that the effective shape becomes arch-like and well described by an inverted catenary. The catenary center lines ($x=c$) roughly track the intruders' centers of mass ($x=x_{m}$). The two lines begin to converge as the intruder approaches a more symmetrical shape. Variation in intruder geometry results in scaling and shifting of the catenary and this determines the lift force on the intruder. We also demonstrate that the mechanism for lift in dilute granular flows consists of at least three processes rather than just collisional forces as might be naively expected. 

In dilute flows, the quasi-static granular pack is analogous to a hydrodynamic radius and determines the effective shape of an intruding object. The existence of several measurements to identify the shock boundary and universality in its shape should pave the way for a better understanding of boundary conditions and more refined applications of Navier-Stokes-like continuum models to dilute granular flows. Our work also presents avenues for future exploration of drag forces on intruder shape, size and impact speeds. This future work would further detail the extent to which granular lift is drag-induced.

\bibliography{granularWindTunnelPaper}

\end{document}


\title{Universality in quasi-2D granular shock fronts above an intruder: Supplementary Materials}
\date{\today}
\author{M. Yasinul Karim, Eric I. Corwin}
\affiliation{Materials Science Institute and Department of Physics, University of Oregon, Eugene, Oregon 97403}

\maketitle

\section{Forces on grain pile}
The lift force $F_{L}$ can be described by a simple model consisting of three parts: the horizontal component of the static load $F_{S}$, the horizontal component of the impact forces $F_{C}$, and the horizontal forces due to momentum transfer of outflowing particles $F_{flow}$.  

The static pile above the intruder has a mass proportional to the volume enclosed by the shock front and the intruder profile, shown by the grey region in Figure \ref{diagram}. For a given exponent $n$, the shape of the intruder profile is given by $g(x)=(1-x^n)^{1/n}$ and the shock front $f(x)$ given by Equation 5 in the main article. For a small segment of of the trapped area of width $dx$ and thickness $w_{cell}$ the volume $dV=w_{cell}[f(x)-g(x)]dx$. The net horizontal component of this weight is given by
\begin{equation} 
F_{S} = \int_0^1  w_{cell}  [f(x)-g(x)] \phi\rho_{g}g\cos(\alpha(x)) \sin(\alpha(x))dx
\label{fstatic}
\end{equation}
where the density of glass is $\rho_{g}=2500kg/m^3$, $g$ is gravitational acceleration and $\alpha(x)$ the angle between the horizontal and tangent to $g(x)$. We approximate the volume fraction $\phi=0.6$, slightly lower than random close packing because of the presence of confining walls in our system which is about 8 bead diameters thick. The cosine term gives the normal component to the tangent and the sine resolves the horizontal component. The horizontal component of the force is integrated over the length of the intruder to calculate the lift force due to the static pack. The integration limits are $x\in[0,1]$ where $x=0$ and $x=1$ are the left and right edges of the intruder respectively.

\begin{figure}[h!]
	\includegraphics[width=\linewidth]{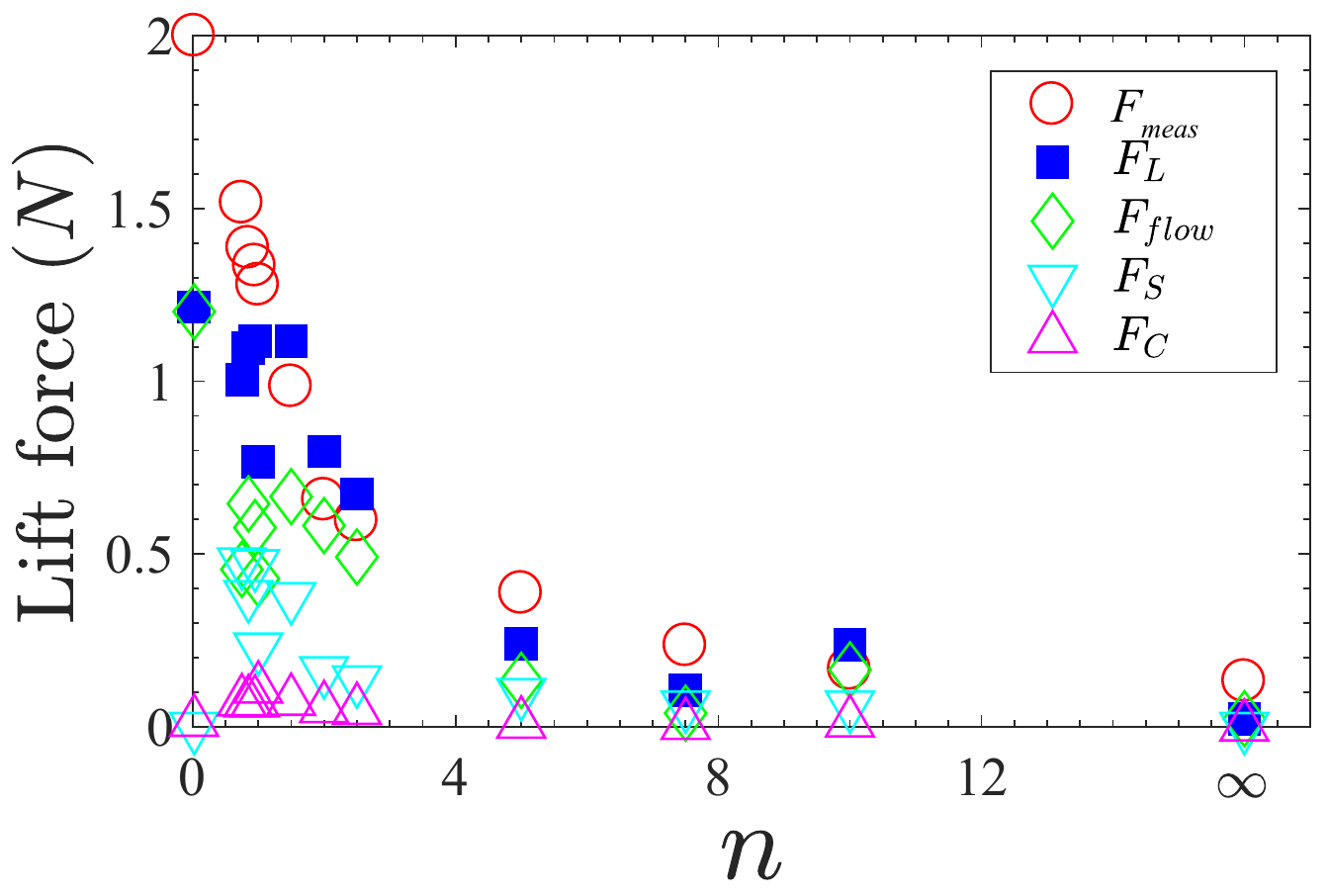}
	\caption{Plots of total measured lift force $F_{meas}$, calculated lift force $F_L$ (blue squares), force due to mass ejection from trapped pile $F_{flow}$ (green diamond), quasi-static load contribution $F_S$ (downward pointing arrow), and force due to impacts $F_C$ (upward pointing arrow) as a function of $n$.}
	\label{forcePlotSupplement}
\end{figure}

The collisional force $F_{C}$ due to beads impacting the shock boundary and is given by 
\begin{equation} 
F_{C} = \int_0^1 \phi_{d}\rho_{g}w_{cell} [\vec{v}(x)\cdot \hat{n}(x)]^{2} \sin{\theta(x)}dx
\label{fcollision}
\end{equation}
where $\phi_{d}\rho_{g}w_{cell} \vec{v}(x)\cdot \hat{n}(x)dx$ is the incident mass per unit time at $x$ on a small segment $dx$ of the shock front. $\theta(x)$ is the angle between the horizontal and tangent to $f(x)$.  The normal component of the incident velocity of particles colliding with the shock front is given by $\vec{v}(x)\cdot \hat{n}(x)$. The fraction of space occupied by freely falling particles is $\phi_{d}\approx0.03$ as measured directly from the images. Assuming that collisions are inelastic, the rate of momentum transfer normal to $f(x)$ is $\phi_{d}\rho_{g}w_{cell} \left(\vec{v}(x)\cdot \hat{n}(x)dx \right) \left(\vec{v}(x)\cdot \hat{n}(x)\right)\sin\theta(x)$. The sine term comes from the horizontal component of this force at $g(x)$.

\begin{figure}[h!]
	\includegraphics[width=\linewidth]{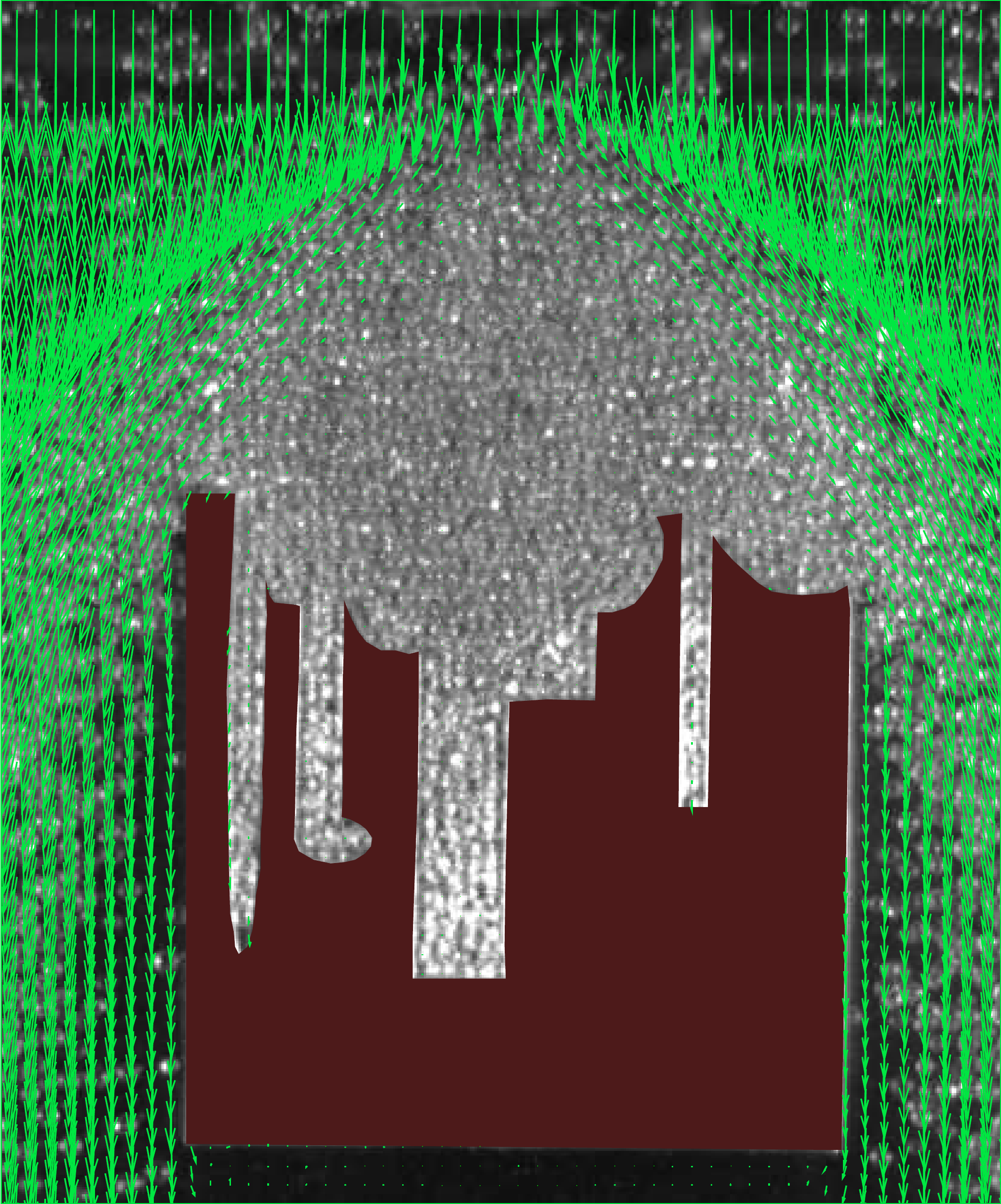}
	\caption{Representative mean flow field around an intruder with super-disk exponent $n=0.75$. The white masked out region is the intruder and the arrows are velocity vectors.}
	\label{flowField}
\end{figure}

The horizontal reaction force $F_{flow}$ on the intruder due to mass ejection from the granular pile  is 
\begin{equation} 
F_{flow} =  \phi \rho w_{cell} \left( \left[ f(1)-g(1) \right] v_{x,1}^{2} - \left[ f(0)-g(0) \right] v_{x,0}^{2} \right)
\label{fflow}
\end{equation}
where $\phi \rho w_{cell}([f(x_0)-g(x_0)] v^{2}$ is the horizontal momentum transferred due to particles being ejected from the area between $f(x_0)$ and $g(x_0))$. The velocities $v_{x,0}$ and $v_{x,1}$ are the mean horizontal bead velocities exiting the cross-section $w_{cell}[f(x)-g(x)]$ at $x=0$ and $x=1$ respectively. The volume fraction is $\phi=0.6$.